\documentclass[fleqn,usenatbib]{mnras}

\usepackage{newtxtext,newtxmath}

\usepackage[T1]{fontenc}
\usepackage{ae,aecompl}

\usepackage{graphicx}	
\usepackage{amsmath}	
\usepackage{amssymb}	
\usepackage{hyperref}

\title[Measuring $\rm ^{12}C/^{13}C$ with ESPRESSO]{A bound on the $\rm ^{12}C/^{13}C$ ratio in near-pristine gas with ESPRESSO.\thanks{Based on observations collected at the European Organisation for Astronomical Research in the Southern Hemisphere, Chile [VLT program IDs: 60.A-9508(A), 086.A-0204(A)], and the W. M. Keck Observatory [Keck program ID: A185Hb], which is operated as a scientific partnership among the California Institute of Technology, the University of California and NASA, and was made possible by the generous financial support of the W. M. Keck Foundation. We also utilise observations collected with the William Herschel Telescope [ING program ID: W/2019B/4] operated on the island of La Palma by the Isaac Newton Group of Telescopes in the Spanish Observatorio del Roque de los Muchachos of the Instituto de Astrof\'isica de Canarias.}}

\author[L. Welsh et al.]{
Louise Welsh,$^{1}$\thanks{E-mail: louise.a.welsh@durham.ac.uk}
Ryan Cooke,$^{1}$
Michele Fumagalli$^{1,2,3}$
Max Pettini$^{4}$
\\
$^{1}$Centre for Extragalactic Astronomy, Durham University, South Road, Durham DH1 3LE, UK \\
$^{2}$Institute for Computational Cosmology, Durham University, South Road, Durham DH1 3LE, UK \\
$^{3}$Dipartimento di Fisica G. Occhialini, Universit\`a degli Studi di Milano Bicocca, Piazza della Scienza 3, 20126 Milano, Italy\\
$^{4}$Institute of Astronomy, University of Cambridge, Madingley Road, Cambridge CB3 0HA, UK \\}

\date{Accepted XXX. Received YYY; in original form ZZZ}

\pubyear{2020}

\usepackage[usenames,dvipsnames]{color}

\defcitealias{CampbellLattanzio2008}{CL08}
\defcitealias{Karakas2010}{K10}
\defcitealias{HegerWoosley2010}{HW10}

\begin{document}
\label{firstpage}
\pagerange{\pageref{firstpage}--\pageref{lastpage}}
\maketitle

\begin{abstract}
Using science verification observations obtained with ESPRESSO at the Very Large Telescope (VLT) in 4UT mode, we report the first bound on the carbon isotope ratio $\rm ^{12}C/^{13}C$ of a quiescent, near-pristine damped Ly$\alpha$ (DLA) system at $z=2.34$. We recover a limit $\rm log_{10}\,^{12}C/^{13}C > +0.37\, (2\sigma)$. We use the abundance pattern of this DLA, combined with a stochastic chemical enrichment model, to infer the properties of the enriching stars, finding the total gas mass of this system to be $\log_{10}(M_{\rm gas}/{\rm M_{\odot}})=6.3^{+1.4}_{-0.9}$ and the total stellar mass to be $\log_{10}(M_{\star}/{\rm M_{\odot}})=4.8\pm 1.3$. The current observations disfavour enrichment by metal-poor Asymptotic Giant Branch (AGB) stars with masses $\rm <2.4\,M_{\odot}$, limiting the epoch at which this DLA formed most of its enriching stars. Our modelling suggests that this DLA formed very few stars until $\gtrsim1$~Gyr after the cosmic reionization of hydrogen and, despite its very low metallicity ($\sim1/1000$ of solar), this DLA appears to have formed most of its stars in the past few hundred Myr. Combining the inferred star formation history with evidence that some of the \emph{most} metal-poor DLAs display an elevated [C/O] ratio at redshift $z \lesssim 3$, we suggest that very metal-poor DLAs may have been affected by reionization quenching. Finally, given the simplicity and quiescence of the absorption features associated with the DLA studied here, we use these ESPRESSO data to place a bound on the possible variability of the fine-structure constant, $\Delta\alpha/\alpha=(-1.2 \pm 1.1)\times10^{-5}$.
\end{abstract}




\section{Introduction}
\label{sec:intro}
\indent The earliest episodes of star formation can be studied by measuring the chemical composition of near-pristine environments. Indeed, there may be some environments in the Universe that have been solely enriched by the first generation of metal-free stars (also known as Population III stars) --- a population of stars that we still know very little about; we are yet to discover a star that shows no detectable metals. However, dedicated surveys (e.g. \citealt{Bond1980, Beers1985, Beers1992, McWilliam1995, Cayrel2004, Beers2008, Howes2016}) have revealed an interesting trend in the chemical composition of the  lowest metallicity stars. Notably, there is an overabundance of carbon in some of the most iron-poor stars found in the halo of the Milky Way; indeed, every star with a measured iron abundance [Fe/H]~$\leq -5.0$ exhibits a strong carbon enhancement\footnote{Here, and throughout this paper,  [X/Y] denotes the logarithmic number abundance ratio of elements X and Y relative to their solar values X$_{\odot}$ and Y$_{\odot}$, i.e. $[{\rm X / Y}] =  \log_{10}\big( N_{{\rm X}}/N_{{\rm Y}}\big) - \log_{10} \big(N_{{\rm X}}/N_{{\rm Y}}\big)_{\odot}$.}  \citep{Christlieb2004,Frebel2005, Aoki2006,Frebel2015,AlPrieto2015, Nordlander2019}. \\
\indent Despite concentrated efforts, and increasingly sophisticated cosmological hydrodynamic simulations, we have yet to establish whether or not low mass ($\rm\sim 1\,M_{\odot}$) metal-free stars can form. Seminal simulations of Population III star formation, like those of \citet{Tegmark1997, BarkanaLoeb2001, Abel2002, BrommCoppiLarson2002}, suggested an initial mass range from $100-1000$~M$_{\odot}$. As the resolution of these simulations improved, alongside our ability to incorporate more detailed physics, the predicted minimum mass of the first stars has decreased. Current simulations suggest that Population III stars were dominated by stars in the mass range $10-100$~M$_{\odot}$ \citep{Turk2009,Greif2010,Clark2011,Hirano2014,Stacy2016}. Given that we are yet to discover a metal-free star around the Milky Way, one might conclude that Population III stars were dominated by more massive  ($>10$~M$_{\odot}$) stars that lived relatively short lives. There are, however, simulations that suggest low mass Population III stars can form, either through efficient fragmentation of the primordial gas \citep{Clark2011, Stacy2016} or through the re-cooling of preserved pristine gas that has been photoionised by a nearby burst of metal-free star formation \citep{Stacy2014}. \\
\indent A complementary approach to studying the imprints of Population III stars in second generation (Population II) stars is the analysis of metal-poor absorption line systems \citep{Erni2006, Pettini2008, Penprase2010}. Of all known damped Ly$\alpha$ systems (DLAs, which are defined as absorption line systems with a neutral hydrogen column density $\log_{10}\,N(\rm H\,${\sc i}$)/\rm cm^{-2}>20.3$), only one gas cloud reportedly shows a carbon enhancement similar to that seen in metal-poor halo stars (\citealt{Cooke2011a}; see also, \citealt{Dutta2014, CookePettiniJorgenson2015}). This system is located at a redshift $z_{\rm abs}\simeq2.34$ along the line-of-sight to the quasar SDSS J003501.88$-$091817.6 (hereafter J0035$-$0918), and displays a large column density of neutral hydrogen, $\log_{10}\,N(\rm H\,${\sc i}$)/\rm cm^{-2}=20.43 \pm 0.04$.  Previous observations of this DLA towards J0035$-$0918 have shown that it is one of the least polluted reservoirs of neutral gas currently known, with a relative iron abundance almost $1/1000$ that of the Sun. DLAs are thought to be self-shielded from external radiation due to their large H\,{\sc i} column density; their constituent elements tend to exist in a single, dominant ionisation state. We can therefore determine the chemical abundance patterns of these systems without needing to apply ionisation corrections. We note there are some reservoirs of partially ionised gas that show \emph{no} detectable metals (e.g. \citealt{Fumagalli2012} and \citealt{Robert2019}). The metal paucity of the DLA towards J0035$-$0918, alongside the originally reported overabundance of carbon, makes this an interesting environment to search for signatures of Population III stars. \\
\indent Here, we propose an observational approach to assess the existence or absence of low mass Population III stars. Simulations of stellar evolution have shown that most stellar populations predominantly produce $^{12}$C. There are only two channels through which low values of $\rm ^{12}C/^{13}C$ can be produced. These involve either: (1) low mass metal-free stars; or (2) metal-poor intermediate mass Asymptotic Giant Branch (AGB) stars \citep{CampbellLattanzio2008, Karakas2010}. Note that both metal-free and metal-enriched massive stars (i.e. $M~>10$~M$_{\odot}$) produce $\rm ^{12}C/^{13}C~>100$ \citep{HegerWoosley2010}. Therefore, by measuring the carbon isotope ratio of a near-pristine gas cloud, we can test if low mass Population III stars might have contributed to the enrichment. Moreover, because stars of different mass produce different quantities of  $\rm ^{12}C/^{13}C$, we can use the measured abundance as a `clock' to infer the enrichment time scale of a system. As only the intermediate mass metal-poor stars produce significant yields of $\rm ^{13}C$, there is a finite time in which a system will retain a distinctive low $\rm ^{12}C/^{13}C$ signature before the $^{12}$C-rich yields of low mass stars return the isotope ratio to larger values. \\ 
\indent A measurement of the carbon isotope ratio in near-pristine gas relies on our ability to distinguish absorption lines that are separated by a small isotope shift; for the C\,{\sc ii}\,$\lambda$1334 absorption line, $^{13}$C is shifted by $\rm -2.99~km~s^{-1}$ relative to $^{12}$C. In typical metal-poor DLAs, the overall line profile contains a small number ($\lesssim5$) of absorption clouds spread over a velocity interval of tens km~s$^{-1}$, where each individual absorption component exhibits a total line broadening of $3-5~{\rm km~s}^{-1}$. The DLA towards J0035$-$0918 is particularly quiescent, where the absorption is concentrated in a single component with an estimated total Doppler width of $b \simeq 3.5~{\rm km~s}^{-1}$ \citep{CookePettiniJorgenson2015}, which is related to the velocity dispersion of the gas, $\sigma=\sqrt{2}\,b$. With such a system, it may be possible to detect $^{13}$C as an asymmetry of the C\,{\sc ii}\,$\lambda$1334 feature. Such an asymmetry will not be present in the absorption lines of other elements. To reach the required level of accuracy, we need to employ a very high spectral resolution instrument that has an accurate wavelength calibration. Such a requirement is now met with the Echelle SPectrograph for Rocky Exoplanet and Stable Spectroscopic Observations (ESPRESSO; \citealt{Pepe2010}) at the European Southern Observatory (ESO) Very Large Telescope (VLT). This high resolution ($\rm R~\simeq~70,000-140,000$) spectrograph provides an unprecedented level of wavelength accuracy; when used in 4UT mode, the relative velocity accuracy is better than $\rm~5~m~s^{-1}$, corresponding to an accuracy of $\sim10^{-4}$~\AA\, at $4000$~\AA. \\
\indent In this paper, we present the first bound on the $\rm ^{12}C/^{13}C$ isotope ratio of the DLA towards J0035$-$0918 using ESPRESSO data obtained during the science verification process. These data have also allowed us to explore the chemical enrichment history of this DLA and to place a bound on the fine-structure constant variation. The paper is organised as follows. Section \ref{sec:obs} describes our observations and data reduction. In Section \ref{sec:results} we present our data and determine the chemical composition of the DLA towards J0035$-$0918 using the detected metal absorption line profiles. We then discuss the chemical enrichment history of this system and infer some of its physical properties in Section \ref{sec:ana}, before drawing overall conclusions and suggesting future work in Section \ref{sec:con}.\\

\section{Observations and data reduction}
\label{sec:obs}
J0035$-$0918 is a $m_{r} = 18.89$ quasar at $z_{\rm em}=2.42$ whose line-of-sight intersects a large pocket of neutral hydrogen. This intervening gas cloud was identified as a DLA at $z_{\rm abs} \simeq 2.34$ from the Sloan Digital Sky Survey (SDSS) discovery spectrum. The lack of metal lines associated with this DLA motivated follow up observations 
with the HIgh Resolution Echelle Spectrograph (HIRES; \citealt{Vogt1994}) on the Keck I telescope by \cite{Cooke2011a}. Further observations were carried out using the Ultraviolet and Visual Echelle Spectrograph (UVES; \citealt{Dekker2000}) on the VLT by \cite{Dutta2014}. The combined analysis of these UVES and HIRES data revealed that the DLA towards J0035$-$0918 is one of the least polluted neutral gas clouds currently known with [Fe/H]~$=-2.94 \pm 0.06$ \citep{CookePettiniJorgenson2015}. 
Furthermore, it is the only very metal-poor DLA to show an overabundance of carbon relative to iron, [C/Fe]~$= +0.58 \pm 0.16$ \citep{CookePettiniJorgenson2015}. 
An enhancement of carbon is thought to be a chemical signature of Population III enriched systems \citep{Beers2005}.
While the DLA towards J0035$-$0918 is not carbon-enhanced to the same degree exhibited by some metal-poor stars, its enhancement of C and N relative to Fe is a rarity amongst the very metal-poor DLA population. \\ 
\indent Previous observations indicated that the DLA towards J0035$-$0918 is particularly quiescent, with a single absorption component exhibiting a total Doppler parameter of $b\simeq 3.5$~km~s$^{-1}$ \citep{CookePettiniJorgenson2015}. The $^{12}$C and $^{13}$C isotopes produce rest frame absorption features at $\lambda$1334.5323~\AA\, and $\lambda$1334.519~\AA\,, respectively \citep{Morton2003}; the isotope shift is therefore just 2.99~km~s$^{-1}$. Given the narrow broadening of this system, it is therefore possible to distinguish between the contribution of each C isotope to the total C\,{\sc ii}\,$\lambda$1334 line profile. Thus, the DLA towards J0035$-$0918 is a near-ideal environment to measure the carbon isotope ratio and search for the chemical signature of low mass ($\rm \sim1~M_{\odot}$) Population III stars. Given the potential promise of this system, we secured new observations with the ultra-stable ESO ESPRESSO spectrograph. \\
\indent New data were collected with ESPRESSO in 4UT mode (R~$\simeq~70,000$) on 2019 August 28 spanning the wavelength range 3800 to 7880\AA.
We acquired 3$\times$2100~s exposures on target using $8\times4$ binning in slow readout mode. In 4UT mode, the light from the four UTs is incoherently sent to ESPRESSO. The size of the entrance fibre at each UT is $1.0$~arcsec. The average seeing during our observations was $0.64$~arcsec. These data were reduced using the EsoRex pipeline, including the standard reduction steps of subtracting the detector bias, locating and tracing the echelle orders, flat-fielding, extracting the 1D spectrum, performing a wavelength calibration, and relative flux calibrations. \\
\indent Due to the faint nature of our target, we have not performed the conventional sky subtraction which would introduce additional sky and read noise into our data.\footnote{This is because the science and sky fibres project to the same number of pixels on the detector. For faint objects, performing a sky subtraction results in counting the sky and read noise twice.} Given that we are already nearing the magnitude limit of what is feasible with this instrument, we decided to maximise the final combined signal-to-noise ratio (S/N) and model the zero level of the data in our analysis. This will be discussed further in Section \ref{sec:results}. \\
\indent As a final step, we combined the three individual exposures using {\sc uves\textunderscore popler}\footnote{{\sc uves\textunderscore popler} is available from:\\ \url{https://astronomy.swin.edu.au/~mmurphy/UVES_popler/}} with a pixel sampling of 2~km~s$^{-1}$. {\sc uves\textunderscore popler} allowed us to manually mask cosmic rays and minor defects from the combined spectrum. The final combined S/N of the data near the  C\,{\sc ii}\,$\lambda$1334 absorption line (at observed wavelength 
$\lambda_{\rm obs} = 4457.8$\,\AA) is S/N~$\simeq$~9. The peak S/N of the data is near 5300~\AA\, (S/N~$\simeq$~30). As will be described in the following section, alongside these ESPRESSO data, we utilise the spectra from previous observations of the DLA towards J0035$-$0918 to model the metal line profiles of this absorption system. These data were recorded with a resolution of $R\simeq40,000$ and a reported S/N per pixel of S/N~$\simeq18$ at 4500~\AA\, \citep{Cooke2011a} and S/N~$\simeq13$ at 5000~\AA\, \citep{Dutta2014}. We refer the reader to these publications for details of these data.
\begin{figure*}
    \centering
    \includegraphics[width=\textwidth]{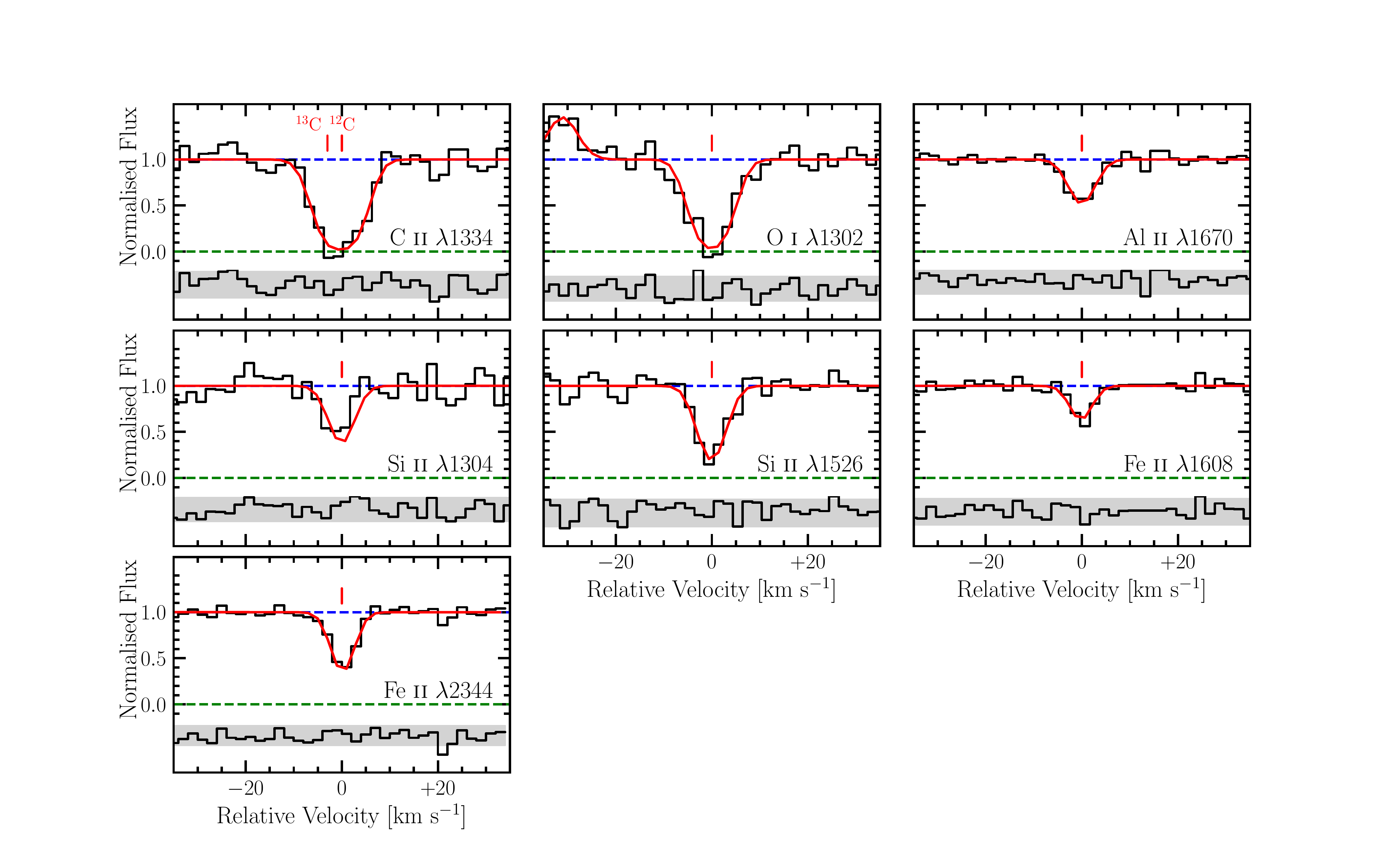}
    \caption{Continuum normalised ESPRESSO data (black histograms) of the absorption features produced by metal ions associated with the DLA at $z_{\rm abs}=2.340064$ towards the quasar J0035$-$0918. Overplotted in red is our best-fitting model. The blue dashed line indicates the position of the continuum while the green dashed line indicates the zero-level. The red ticks above the absorption features indicate the centre of the Voigt line profiles. In the panel that shows C~{\sc ii} $\lambda$1334 absorption, the tick marks at a relative velocity of $0.0~\rm km~s^{-1}$ and $-2.99~\rm km~s^{-1}$ represent the centroid of the $^{12}$C and $^{13}$C absorption line profiles, respectively. Below the zero-level, we show the residuals of this fit (black histogram) where the grey shaded band encompasses the $2\sigma$ deviates between the model and the data. Note, in the panel corresponding to O\,{\sc i}~$\lambda1302$, there is an unrelated emission line (due to an unrelated, intervening galaxy at redshift $z\simeq0.15$; see text) that we have modelled as a Gaussian during the line-fitting procedure.}
    \label{fig:data}
\end{figure*}
\section{Results}
\label{sec:results}
We exploit the superior wavelength calibration of ESPRESSO to pin down the redshift of the DLA using O, Al, Si, and Fe absorption lines, and search for a shift/asymmetry of the C\,{\sc ii} line profile  --- \emph{relative} to the other metals --- indicative of absorption from $\rm ^{13}C$. \\
\indent Using the Absorption LIne Software ({\sc alis}) package\footnote{{\sc alis} is available from:\\ \url{https://github.com/rcooke-ast/ALIS}.}--- which uses a $\chi$-squared minimisation procedure to find the model parameters that best describe the input data --- we simultaneously analyse the full complement of high S/N and high spectral resolution data currently available. While the ESPRESSO data provide the most reliable wavelength solution, the UVES and HIRES data can be leveraged alongside the ESPRESSO data to enable a more accurate determination of the metal ion column densities and assist in the determination of the zero-level of the ESPRESSO data. To achieve this, the redshift of the DLA is driven by the accurate wavelength solution provided by the ESPRESSO data. The centre of each absorption feature in the UVES and HIRES data is then modelled with an independent velocity offset to ensure that these data are coincident with the well-calibrated ESPRESSO data. \\ 
\indent We model the absorption line profiles as a single component Voigt profile, which consists of three parameters: a column density, a redshift, and a line broadening. We assume that all lines of comparable ionisation level have the same redshift, and any absorption lines that are produced by the same ion all have the same column density. The total broadening of the lines includes a contribution from both turbulent and thermal broadening. The turbulent broadening is assumed to be the same for all absorption features, while the thermal broadening 
depends inversely on the square root of the ion mass; thus, heavy elements (e.g. Fe) will exhibit intrinsically narrower absorption profiles than lighter elements, (e.g. C). 
There is an additional contribution to the line broadening due to the instrument. \\
\indent The nominal ESPRESSO instrument resolution in 4UT mode is $v_{\rm FWHM}=4.28$~km~s$^{-1}$, and we have explicitly checked this by measuring the widths of ThAr emission lines from the calibration data to infer the instrument full width at half maximum (FWHM) at wavelengths close to the DLA's absorption features. We find that across the wavelength range ($4450-7830$)~\AA\, the wavelength specific FWHM varies from 4.14~to~4.53~~km~s$^{-1}$. We adopt these wavelength specific resolutions as our fiducial choice when fitting the data. However, we also checked that our results did not change when using the nominal instrument FWHM. For the HIRES and UVES data, the respective nominal instrument resolutions are $v_{\rm FWHM}= 8.1$~km~s$^{-1}$ and $v_{\rm FWHM} =7.75$~km~s$^{-1}$.  When fitting to the data we allow these to vary as free parameters, since the DLA absorption features are unresolved in the UVES and HIRES data. We found that the choice of instrument resolution does not have a significant impact on the resulting column densities. \\
\indent Finally, we note that we simultaneously fit the absorption, quasar continuum, and in the case of the ESPRESSO data, the zero-level of the data. We model the continuum around every absorption line as a low order Legendre polynomial (of order ~3). We assume that the zero-levels of the sky-subtracted UVES and HIRES data do not depart from zero (this is confirmed by measuring the troughs of saturated absorption lines). Upon inspection of the sky-subtracted ESPRESSO data, we identified a few emission lines that appear to be due to contamination by a nearby galaxy in the field. These emission lines are not present in the UVES or HIRES data. The redshift of this galaxy was determined through observations using the William Herschel Telescope (WHT) that revealed strong O\,{\sc ii}~$\lambda 3727$~\AA\, emission at an observed wavelength $\lambda\simeq4297$~\AA\,, corresponding to $z\simeq0.15$. This galaxy is just $6.8$~arcsec from the line-of-sight to the background quasar (corresponding to an impact parameter of 21~kpc at the redshift of the intervening galaxy). We confirmed that there are no sky or galaxy emission lines that contaminate the DLA absorption lines. Therefore to account for the sky continuum and potential low-level contamination by the continuum of this low redshift galaxy, we include a single parameter to model the zero-level of the ESPRESSO data (assumed constant for all lines). \\

\subsection{Ion column densities}
The data, along with the best-fitting model are presented in Figure \ref{fig:data}, while the corresponding column densities are listed in Table \ref{tab:results}.
\begin{table}
	\centering
	\caption{Ion column densities of the DLA at $z_{\rm abs}=2.340064$ towards the quasar J0035$-$0918. The quoted column densities are based on the combined fit of the available ESPRESSO, UVES and HIRES data. 
	The quoted column density errors are the $1\sigma$ confidence limits.}
	\label{tab:results}
	\tabcolsep=0cm
\begin{tabular}{cc}
\hline
\textbf{Ion}                                                                                                   & \textbf{log$_{10}$ $N$(X)/cm$^{-2}$} \\ \hline \hline
H\,{\sc i}$^{\rm a}$         & 20.43 $\pm$ 0.04                                 \\
$^{12}$C\,{\sc ii} + $^{13}$C\,{\sc ii}                                             & 14.29 $\pm$ 0.13                                  \\
N\,{\sc i}                                                                                                   & 13.37 $\pm$ 0.04                                  \\
O\,{\sc i}                                                                                                   & 14.67 $\pm$ 0.05                                  \\
Mg\,{\sc ii}                                                                                                  & 12.89 $\pm$ 0.13                                  \\
Al\,{\sc ii}                                                                                                 & 11.74 $\pm$ 0.04                                  \\
Si\,{\sc ii}                                                                                                 & 13.35 $\pm$ 0.04                                  \\
Fe\,{\sc ii}                                                                                                 & 13.01 $\pm$ 0.03                                  \\ \hline
\multicolumn{2}{|c|}{\textbf{Isotope ratio}} \\ \hline \hline
\multicolumn{2}{|c|}{$\log_{10}\,\rm ^{12}C/^{13}$C $>+0.37~(2\sigma)$}   \\ \hline
\multicolumn{2}{|c|}{$^{\rm a}$~The H\,\textsc{i} column density was reported by \protect\citet{CookePettiniJorgenson2015}.} 
\end{tabular}
\end{table}
The simultaneous analysis of the ESPRESSO+HIRES+UVES data have allowed us to determine the metal column densities, gas kinetic temperature and total Doppler parameter of the DLA. We find an absorption redshift of $z_{\rm abs}=2.340064\pm 0.000001$ and a gas temperature of $T = (9.1 \pm 0.5)\times 10^{3}$~K. We have found that the line broadening of this system is entirely dominated by its thermal motions; the minor contribution due to the turbulent motions cannot be determined given the current data and, intriguingly, is consistent with no turbulence. The extreme quiescence of this system raises an interesting possibility about the existence of other thermally dominated DLAs yet to be found, or whether known systems that contain multiple absorption components, may still have components that are dominated by thermal broadening. \\
\begin{figure*}
    \centering
    \includegraphics[width=\textwidth]{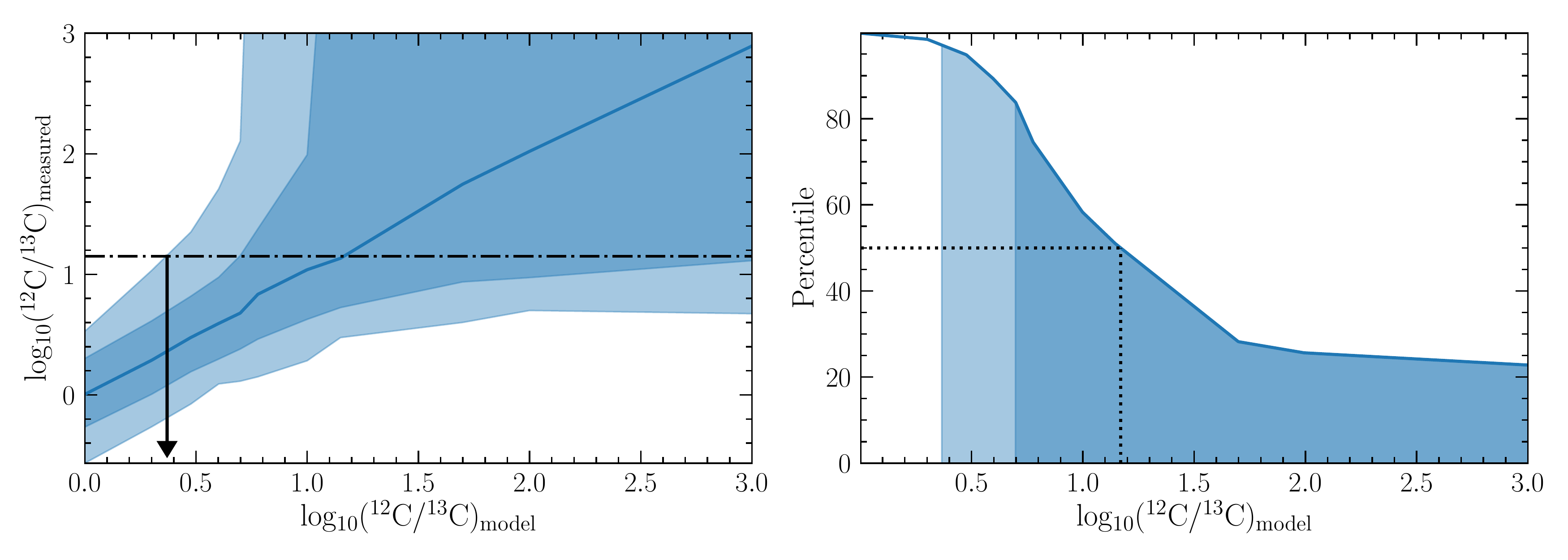}
    \caption{Monte Carlo simulations of our data used to infer a confidence bound on the amount of $^{13}$C in the DLA towards J0035$-$0918 (left panel). The blue line indicates the median recovered value of the $\rm ^{12}C/^{13}C$ ratio given 500 realisations of the absorption feature generated using the model $\rm ^{12}C/^{13}C$ ratio indicated by the x-axis. The dark and light blue shaded bands encompass the $1\sigma$ and $2\sigma$ limits of the distribution, respectively. The horizontal dot-dashed line marks the $\rm ^{12}C/^{13}C$ measured in our analysis. The black arrow indicates where this value intersects the 97.5 percentile of the distribution. This corresponds to a $2\sigma$ lower limit of $\rm ^{12}C/^{13}C>+ 0.37$. The right panel shows the percentile value as a function of the model (i.e. true) $\rm^{12}C/^{13}C$ isotope ratio given our measured value. The shaded bands have the same meaning as in the left panel. The dotted lines mark the 50$^{\rm th}$ percentile and the corresponding model value.}
    \label{fig:lower_bound}
\end{figure*}
\begin{figure}
    \centering
    \includegraphics[width=\columnwidth]{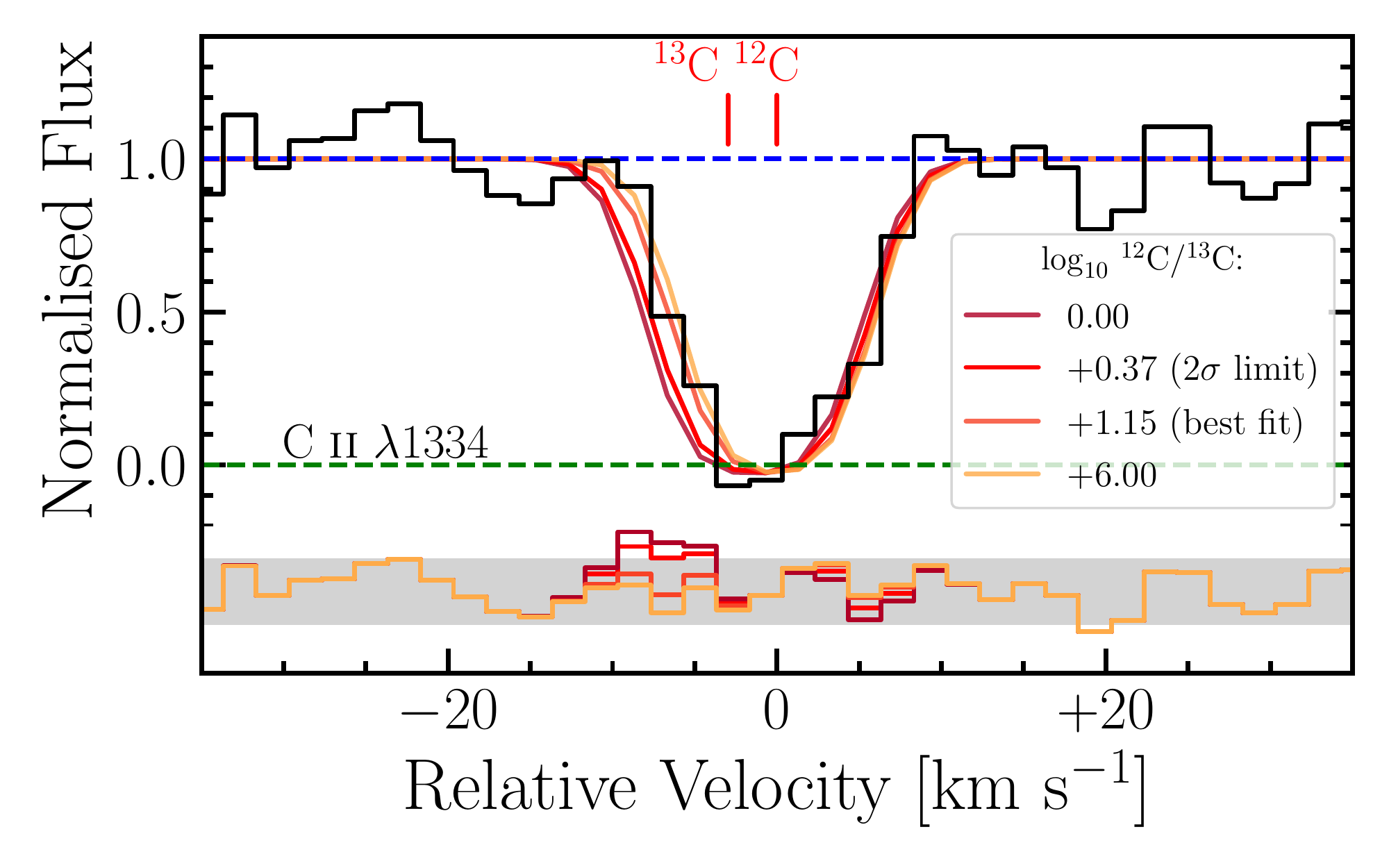}
    \caption{ESPRESSO data centred on the C~{\sc ii} $\lambda$1334 absorption feature shown alongside different model line profiles. Each model curve represents a different $\rm ^{12}C/^{13}C$ abundance ratio (as indicated by the legend) while retaining a constant total carbon abundance of $\log_{10} N({\rm C_{tot})/cm}^{-2} = 14.29$. Below the zero-level of these data (green dashed line), we show the residuals of the model fit to the data. The shaded band encompasses the $2\sigma$ deviations of these model profiles, illustrating that we can rule out $\rm \log_{10}\, ^{12}C/^{13}C \leq +0.37$ with 95 per cent confidence. }
    \label{fig:spec_lower_bound}
\end{figure}
\indent For the C\,{\sc ii} absorption, we found that the best-determined parameter combination was the isotope ratio $\rm ^{12}C/^{13}C$ and the total column density of C\,{\sc ii}, $N(^{12}$C\,{\sc ii})+$N(^{13}$C\,{\sc ii}). The carbon isotope ratio  of this model is $\rm \log_{10}\,^{12}C/^{13}C = +1.15\pm 0.65$, where the quoted error is simply the diagonal term of the covariance matrix. However, given the large range allowed by this uncertainty, we have performed a suite of detailed Monte Carlo simulations to uncover the posterior distribution of the $^{12}$C/$^{13}$C ratio, given our data.
Using the parameters of our best-fitting line model, we generate mock ESPRESSO data varying the relative amount of $^{13}$C in the system. These mock data provide perfect (error free) line profiles of C\,{\sc ii}\,$\lambda1334$ for different values of the $\rm ^{12}C/^{13}C$ isotope ratio (while the total C\,{\sc ii} column density remains constant). By perturbing these line profiles using the error spectrum of our data, we can emulate how these ESPRESSO data would look as a function of the underlying isotope ratio. We have performed 500 realisations of these perturbations for a variety of underlying isotope ratios. The results of these simulations are presented in Figure \ref{fig:lower_bound}. Given that our line fitting procedure, applied to the real data, suggests a central value $\rm \log_{10}\,^{12}C/^{13}C=+1.15$, we infer  $\rm log_{10}\, ^{12}C/^{13}C > +0.37~(2\sigma)$. This lower bound is visualised in Figure \ref{fig:spec_lower_bound}, which shows the model line profiles for various $\rm ^{12}C/^{13}C$ abundance ratios. The line corresponding to $\rm log_{10}\, ^{12}C/^{13}C = +0.37$ falls at the edge of the asymmetric line profile where the $^{13}$C absorption is most noticeable; the corresponding residuals are also at the $2\sigma$ boundary of the model fit. 
As the amount of $^{12}\rm C$ relative to $^{13}\rm C$ in a system increases, the asymmetry due to the presence of $^{13}\rm C$ becomes increasingly subtle in the C line profiles. This is, in part, why we expect to recover a broad range of isotope ratios once  $\rm log_{10}\, ^{12}C/^{13}C > +1.10$. However, we expect with higher S/N data, the threshold for a detection would extend to larger isotope ratios.  \\
\indent This is the first limit on the carbon isotope ratio in a near-pristine absorption system. With these data we can empirically rule out the presence of large amounts of $^{13}$C relative to $^{12}$C in the DLA towards J0035$-$0918. The implications of this abundance ratio for the chemical enrichment of this DLA will be discussed in Section \ref{sec:ana}. The relative abundances of the detected metals are provided for convenience in Table \ref{tab:rel_abuns}. We note that with the latest data, [C/Fe]~=~$+0.32\pm0. 16$. While consistent with the previous determinations by \citet{Carswell2012, Dutta2014} and \citet{CookePettiniJorgenson2015}, this indicates that the DLA towards J0035$-$0918 is not as abundant in carbon as previously thought, owing to the unusual quiescence of the gas cloud whose broadening is dominated by the thermal motions. However, this system still exhibits an unusually high [N/O] ratio, compared with the typical very metal-poor DLA population \citep{Cooke2011b}. The [N/O] abundance of this DLA places it just above the primary N plateau \citep{Pettini2008, Petitjean2008, Zafar2014}. Furthermore, the [Mg/Si], [Mg/O], and [Mg/Fe] ratios are remarkably subsolar, quite unlike the ratios that are seen in extremely metal-poor halo stars of the Milky Way (e.g. \citealt{Andrievsky2010}).\\
\begin{table}
\centering
	\caption{Relative abundances of the elements detected in the DLA towards J0035$-$0918 alongside their solar abundances as determined by \protect\citet{Asplund}. }
	\label{tab:rel_abuns}
\begin{tabular}{ccccc}
\hline
\textbf{X}  & \textbf{[X/H]} & \textbf{{[}X/Fe{]}}  & \textbf{{[}X/O{]}}   & \textbf{X$_{\odot}$} \\ \hline
C  & $-$2.57 $\pm$ 0.14 & $+$0.32 $\pm$ 0.13      & $-$0.12 $\pm$ 0.14     & 8.43                 \\
N  & $-$2.89 $\pm$ 0.06 & \,0.00 $\pm$ 0.05     & $-$0.44 $\pm$ 0.06     & 7.83                 \\
O  & $-$2.45 $\pm$ 0.06 & $+$0.44 $\pm$ 0.06      & \dots & 8.69                 \\
Mg & $-$3.10 $\pm$ 0.14 & $-$0.21 $\pm$ 0.13     & $-$0.65 $\pm$ 0.14     & 7.56                 \\
Al & $-$3.13 $\pm$ 0.06 & $-$0.24 $\pm$ 0.05     & $-$0.68 $\pm$ 0.06     & 6.44                 \\
Si & $-$2.59 $\pm$ 0.06 & $+$0.30 $\pm$ 0.05      & $-$0.14 $\pm$ 0.06     & 7.51                 \\
Fe & $-$2.89 $\pm$ 0.05 & \dots & $-$0.44 $\pm$ 0.06     & 7.47                 \\ \hline
\end{tabular}
\end{table}
\section{Analysis}
\label{sec:ana}
Given our robust determination of the chemical abundance pattern of the DLA towards J0035$-$0918, we now investigate the enrichment history of this system. Our lower bound on the $\rm ^{12}C/^{13}C$ isotope ratio indicates that there is at least 2.3 times more $^{12}$C than $^{13}$C in this DLA; this does not empirically rule out enrichment from low mass Population III stars. To test whether the chemical signature of this system is better modelled by Population III or Population II enrichment, we can exploit the stochastic chemical enrichment model developed by \citet{Welsh2019}, alongside the yields from simulations of stellar evolution. The resulting enrichment models can be used to infer the epoch at which this DLA formed its enriching stars as well as its total stellar mass and total gas mass. In addition to investigating the physical and chemical properties of the DLA towards J0035$-$0918, given the simplicity of the absorption line profiles and the reliable wavelength solution delivered by ESPRESSO, we can also use these data to test the invariance of the fine-structure constant.
\begin{figure*}
    \centering
    \includegraphics[width=.9\textwidth]{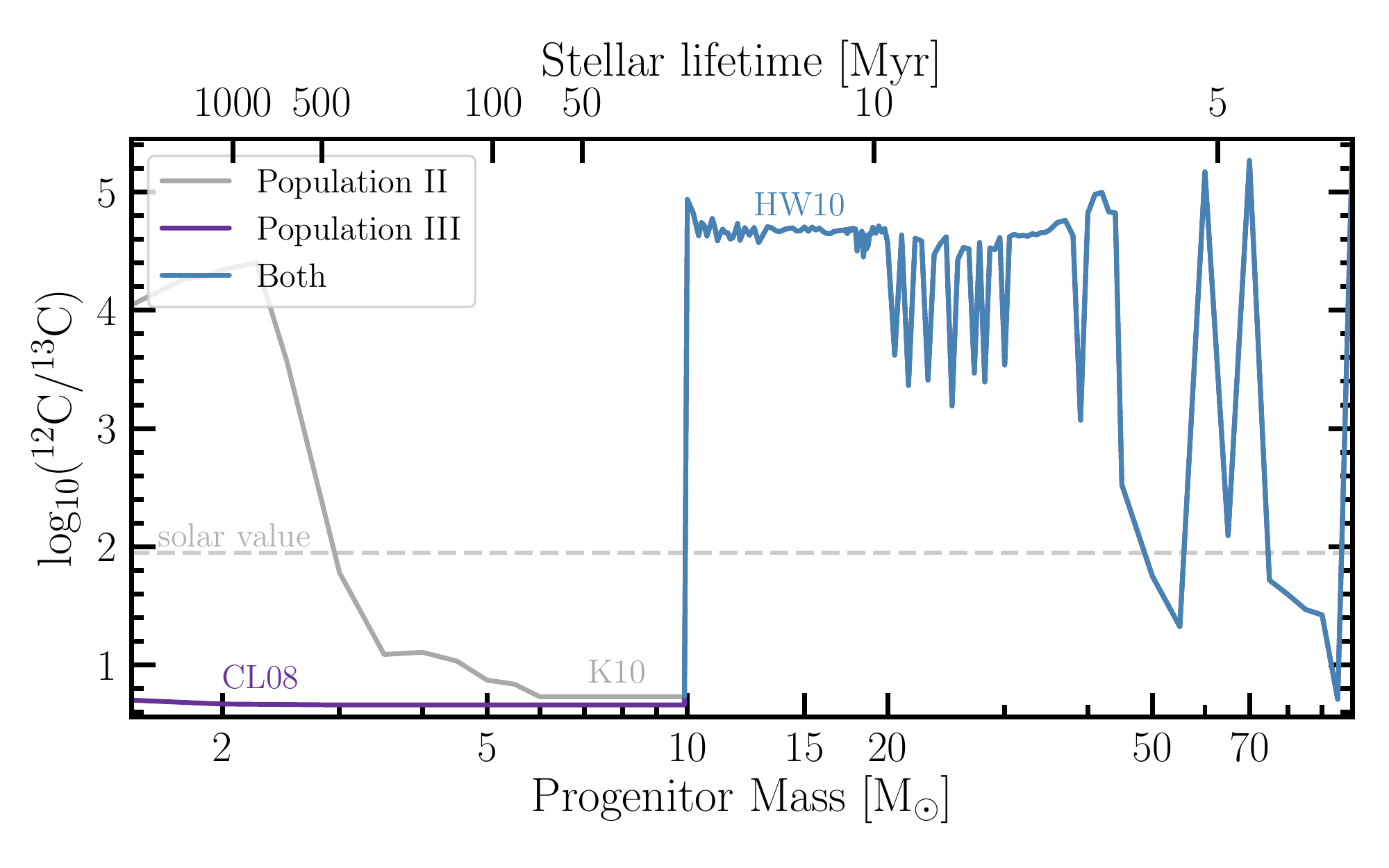}
    \caption{$^{12}$C/$^{13}$C yield as a function of progenitor mass for both Population III and Population II stars. The yields of low mass Population III stars are from \protect\citetalias{CampbellLattanzio2008} while those of Population II stars are from \protect\citetalias{Karakas2010}. The yields of massive Population III and Population II stars are given by \protect\citetalias{HegerWoosley2010}. We have only included the yields of stars whose lifetimes are shorter than the age of the Universe at $z=2.34$ (i.e. those with $M>1.46~{\rm M_{\odot}}$). The stellar lifetimes are indicated by the top x-axis  (\protect\citealt{Woosley2002}, \protect\citealt{Karakas2014}). The horizontal grey dashed line indicates the solar C isotope ratio, $\rm ^{12}C/^{13}C=89$ \protect\citep{Asplund}.}
    \label{fig:C_rat}
\end{figure*}
\subsection{Stochastic Enrichment Model} 
In previous work \citep{Welsh2019}, we developed a stochastic chemical enrichment model that uses the abundance patterns of near-pristine environments to infer their chemical enrichment history. This model describes the initial mass function (IMF) of an enriching stellar population as a power law, governed by the slope, $\alpha$. The normalisation of this power law, $k$, is set by the number of stars, $N_{\star}$, that form within a given mass range: 
\begin{equation}
    \label{eqn:imf}
    N_{\star}  = \int^{M_{\rm max}}_{M_{\rm min}}kM^{-\alpha}dM \;.
\end{equation}
For reference, a Salpeter IMF corresponds to $\alpha = 2.35$ \citep{Salpeter1955}. For a given enrichment model, we then stochastically sample the IMF and, using the yields from simulations of stellar evolution, determine the distribution of chemical abundances we expect to see across an enriched population of objects. These distributions can then be used to gauge the likelihood of measuring our observed abundances given any underlying enrichment model. This approach assumes that the gas within the DLA is well-mixed and that the system experiences no inflow or outflow of gas; for further details of this model, see \citet{Welsh2019}.\\
\indent The stellar yields used in this analysis are from three sources: (1) \citet{CampbellLattanzio2008}, hereafter \citetalias{CampbellLattanzio2008}, who simulate the evolution of low mass metal-free stars in the mass range $(1-3)$~M$_{\odot}$; (2) \citet{Karakas2010}, hereafter \citetalias{Karakas2010}, who simulate the evolution of very metal-poor AGB stars ($Z\sim0.005Z_{\odot}$) in the mass range 
$(1-6)$~M$_{\odot}$; and, (3) \citet{HegerWoosley2010}, hereafter \citetalias{HegerWoosley2010}, who simulate the evolution and core-collapse supernovae (CCSNe) of massive ($>10$~M$_{\odot}$) metal-free stars. Throughout this work, we use the combined yields of \citetalias{CampbellLattanzio2008} and \citetalias{HegerWoosley2010} to define the yields of Population III stars. In lieu of simulations that calculate the $\rm ^{12}C/^{13}C$ ratio for intermediate mass Population III stars, we have chosen to extrapolate the yields of \citetalias{CampbellLattanzio2008} to meet the yields of \citetalias{HegerWoosley2010}.  \\
\indent While the \citetalias{HegerWoosley2010} yields have been calculated for metal-free stars, they are also indicative of Population II CCSNe yields; this can be seen by comparison with the \citet{WoosleyWeaver1995} yields of metal-enriched massive stars (at least for the elements under consideration in this work). We therefore define Population II yields as the combined yields of \citetalias{Karakas2010} and \citetalias{HegerWoosley2010}. We necessarily implement another yield extrapolation to bridge the gap between the \citetalias{Karakas2010} and \citetalias{HegerWoosley2010} yields. Although this extrapolation is not ideal, this is best option available to us until a more complete set of yields becomes available. \citetalias{Karakas2010} report stellar yields covering a range of metallicities, spanning 0.0001$<Z<$0.02. We choose Population II yields with an initial metallicity $Z = 0.0001$. We note that the \citetalias{HegerWoosley2010} yields have been calculated as a function of the progenitor star mass, the explosion energy of their supernova, and the mixing between the different stellar layers. When considering these yields, we have adopted the recommended prescription for mixing between stellar layers, defined to be 10 per cent of the helium core size. For the explosion energy, we adopt $E_{\rm exp} =1.8\times10^{51}$~erg. This is a measure of the final kinetic energy of the ejecta at infinity and is consistent with the typical value found by \citet{Welsh2019} when investigating the properties of the stars that enrich the most metal-poor DLAs. Figure \ref{fig:C_rat} shows the resulting $\rm ^{12}C/^{13}C$ yields of these stellar populations as a function of both their progenitor mass and stellar lifetime. \\
\begin{figure*}
    \centering
    \includegraphics[width=0.8\textwidth]{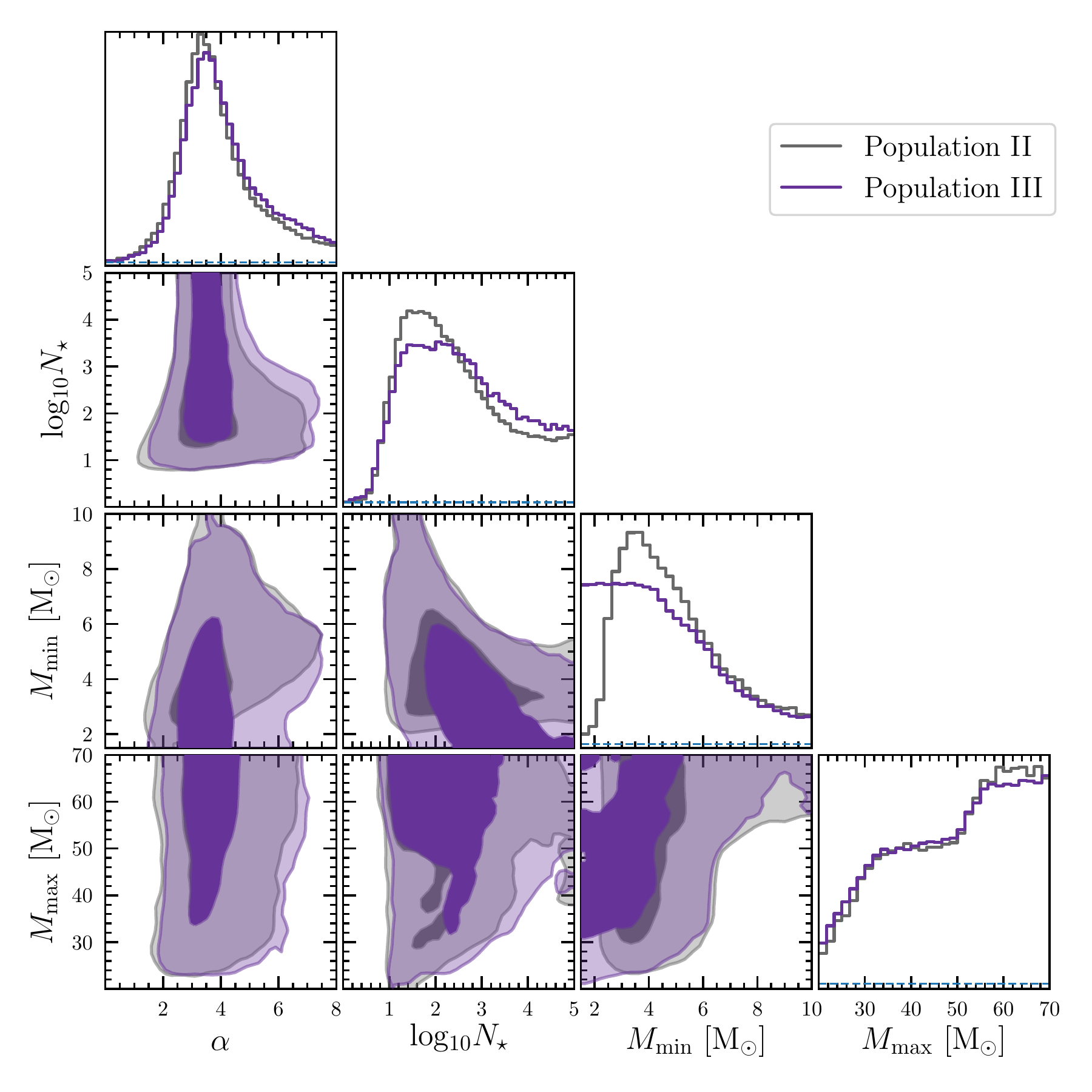}
    \caption{Results of our MCMC analysis of the chemical enrichment of the DLA towards J0035$-$0918. The diagonal panels indicate the maximum likelihood posterior distributions of our enrichment model parameters while the 2D contours indicate the correlation between these parameters. Our Population II model is shown in grey. Overplotted in purple is the result of considering Population III stars as the dominant source of enrichment. In the diagonal panels, the horizontal blue dashed line indicates the zero-level of each distribution.  }
    \label{fig:corner_plot}
\end{figure*}
\indent From Figure \ref{fig:C_rat} we can see that it is the low mass Population III stars, and the intermediate mass Population II stars, that are capable of producing comparable amounts of the carbon isotopes (i.e. $^{12}$C/$^{13}$C$\simeq 1$). Generally, it is surface mixing events that facilitate the production of $^{13}$C. AGB (both Population II and Population III) stars produce $^{13}$C through a process known as hot bottom burning (HBB). This process involves the convection induced transport of $^{12}$C from the burning shell to the proton-rich envelope where $^{13}$C can then be synthesised via proton-capture \citep{Iben1975, Prantzos1996}. HBB is dependent on both the mass and metallicity of the progenitor stars. For a star to undergo HBB, the convective envelope must reach a sufficiently high temperature. 
The transition seen at $\sim10$~M$_{\odot}$ between the yields of massive stars and those of lower mass stars originates because massive stars do not show signs of these surface mixing events \citep{KarakasLattanzio2014}; they are only capable of producing $^{13}$C through secondary processes.  \\
\indent Using these yields alongside our enrichment model, we can investigate the chemical enrichment of the DLA towards J0035$-$0918 under the assumption of either Population II or Population III enrichment. We use the relative abundances of [C/O], [Si/O], and [Fe/O] as given in Table \ref{tab:rel_abuns}, alongside the lower limit on $\rm ^{12}C/^{13}C $ from Table \ref{tab:results} to evaluate the likelihood of a given model. We choose to model only these abundance ratios because our stochastic chemical enrichment model is computationally expensive. We therefore focus our attention on the most abundant elements that are relatively well-modelled by stellar evolution. We use the \textsc{emcee} Markov Chain Monte Carlo (MCMC) sampler \citep{EMCEE} to determine the enrichment model parameters that provide the best fit to these data. The model parameters that we consider are those defined by Equation \ref{eqn:imf} ($\alpha, N_{\star}, M_{\rm min}, M_{\rm max}$). We impose uniform priors on these parameters, limited by the boundary conditions: 
\begin{eqnarray*}
 0 \leq &\log_{10}\,N_{\star}& \leq 5 \; , \\
 1.46\leq &M_{\rm min}/{\rm M_{\odot}}& \leq 11 \; , \\
20\leq &M_{\rm max}/{\rm M_{\odot}}& \leq 70 \; , \\
  -8 \leq &\alpha &\leq 8 \; .
\end{eqnarray*}
\noindent Since the \emph{number} of stars contributing to the enrichment of the DLA, $N_{\star}$, could be quite large in the case of a low value of $M_{\rm min}$, we choose to sample log$_{10}$ N$_{\star}$; this allows us to stochastically sample the IMF at high masses (a regime we suspect may be dominated by just a few massive stars), while still allowing for a much larger number of low mass stars. The minimum mass of the enriching stars is set by considering the age of the Universe at the redshift of the DLA. Using the latest \citet{Planck2018} cosmology, where $H_{0} = 67.4 \pm 0.5$~km~s$^{-1}$Mpc$^{-1}$ and $\Omega_{\rm m} = 0.315 \pm 0.007$, we find that the age of the Universe is 2.792~Gyr at the redshift of this absorption system (recall $z_{\rm abs} = 2.34006$). Given that the first stars are thought to have formed between $z\sim 15-20$, there is a finite time in which stars can contribute to the enrichment of this system. Using the stellar lifetimes from \citet{Karakas2014}, we find that stars with masses $\lesssim 1.46~\rm M_{\odot}$ live longer than the age of the Universe at redshift $z\simeq2.34$. Therefore we only expect to see the chemical signature of stars with masses above this limit in the chemistry of this near-pristine DLA. The upper limit of the mass of the enriching stars marks the transition from CCSNe to pulsational pair instability SNe, as found by \citet{Woosley2017}. \\
\begin{figure*}
    \centering
    \includegraphics[width=\textwidth]{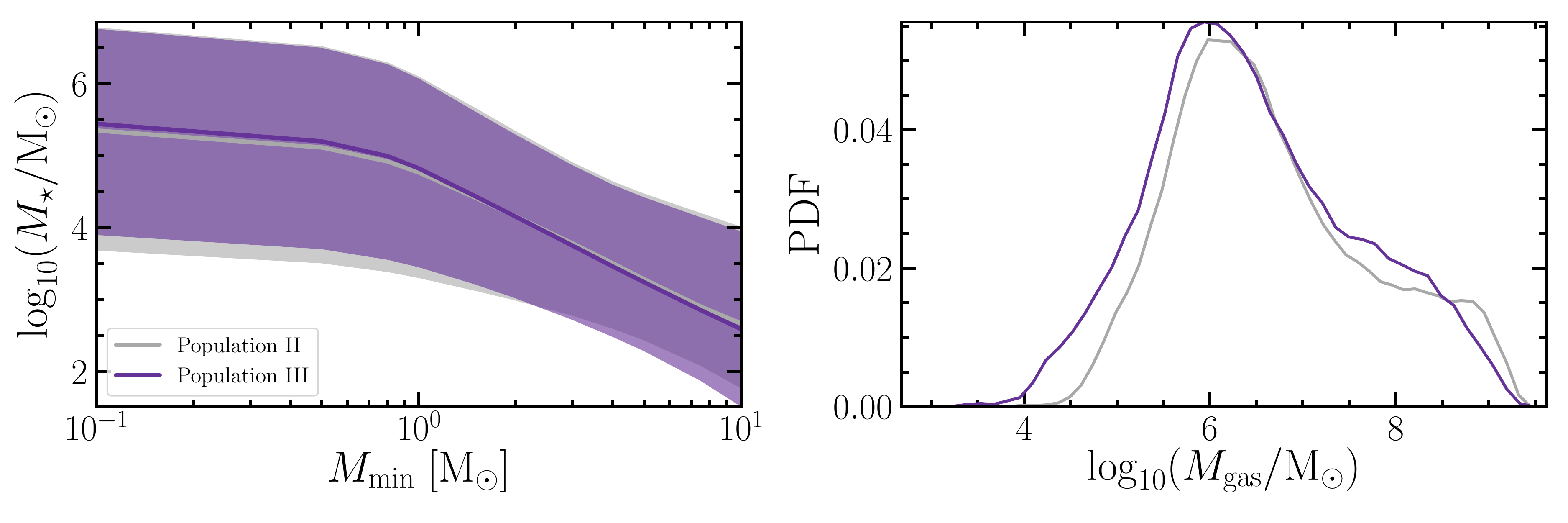}
    \caption{Left: The total inferred stellar mass of the DLA towards J0035$-$0918 as a function of the minimum mass with which stars can form. The purple solid curve indicates the median value given our Population III enrichment model and the shaded region encompasses the 16$^{\rm th}$ and 84$^{\rm th}$ percentiles. The grey curves have the same meaning, but are based on our Population II enrichment model. Right: The total gas mass of the DLA towards J0035$-$0918 inferred from our enrichment models combined with the measured [O/H] abundance ([O/H]$=-2.45$), assuming 100 per cent metal retention (see text for further details).  }
    \label{fig:gas_mass_and_M*}
\end{figure*}
\indent During our MCMC analysis, we utilise the chains of 400 randomly initialised walkers to determine the posterior distributions of our enrichment model parameters. After adopting a burn-in that is half the original length of the chains, we find the posterior distributions shown in Figure \ref{fig:corner_plot}. We found that these distributions are invariant once the walkers have each taken 2100 steps. From this figure, we see that the maximum likelihood enrichment model parameters are almost unchanged by the assumption of Population II versus Population III enrichment. Both enrichment histories suggest an IMF slope that is preferably steeper than, but still consistent with, a Salpeter distribution, $\alpha = 3.6^{+3.7}_{-2.0}$ (Population II) $\alpha = 3.8^{+3.6}_{-2.0}$ (Population III), where the quoted errors encompass 95 per cent of the parameter distributions. We have repeated our analysis under the assumption of a Salpeter-like IMF slope and found that introducing this prior has a negligible impact on the resulting distributions. \\
\indent Our analysis suggests that this DLA has been enriched by at least 10 stars, with maximum likelihood values of $\log_{10} N_{\star} = 2.3^{+2.5}_{-1.4}$ (Population II) and $\log_{10} N_{\star} = 2.5^{+2.3}_{-1.7}$ (Population III). In each enrichment scenario, $M_{\rm max}$ is unconstrained, with both distributions showing a slight preference towards  a larger maximum enriching mass. We find the only parameter estimate that varies significantly between these enrichment histories is that of the minimum enriching mass, $M_{\rm min}$. Under the assumption of Population II enrichment, the data disfavour enrichment from low mass ($<\rm 2.4~M_{\odot}$) stars. While, if this is a Population III enriched system, enrichment from low mass stars is preferable. This is likely being driven by the divergent yields of [C/O] across the stellar populations. From the simulations of stellar evolution, we see that the lowest mass Population II stars produce elevated [C/O] relative to our measured value ([C/O]$~=-0.12\pm0.14$), while the lowest mass Population III stars produce subsolar yields of [C/O]. Given that our maximum likelihood enrichment model is consistent with a well-sampled IMF, to investigate this divergence further, we can calculate the IMF weighted abundance of [C/O] for both Population II and Population III enrichment. These calculations show that, given our maximum likelihood estimate of  $\alpha$, when $M_{\rm min} < 2.4~{\rm M}_{\odot}$ the C-rich yields of the lowest mass Population II stars result in supersolar [C/O]. These yields are  hard to reconcile with our measured value. \\
\indent Given current data, we are only able to utilise the $^{12}$C/$^{13}$C lower bound to constrain our enrichment model. This lower bound does little to drive the results of our current analysis. However, as can be seen from Figure \ref{fig:C_rat}, the C isotope ratio is also divergent at low masses for the different stellar populations. Therefore, a precise measurement of the $\rm ^{12}C/^{13}C$ ratio, in combination with the [C/O] abundance, would enable us to distinguish more clearly whether the DLA towards J0035$-$0918 shows the signature of Population III versus Population II enrichment. \\ 
\subsection{Physical and chemical properties}
\indent Using the results of our enrichment model analysis, we can infer some of the physical and chemical properties of the DLA towards J0035$-$0918, such as the total stellar mass and the total gas mass of the system. To calculate the total stellar mass, we use the inferred model parameter distributions and integrate over the IMF, weighted by mass (cf. Equation \ref{eqn:imf}). For stars above 1~$\rm M_{\odot}$, we adopt a power law IMF, while for stars $<~1~\rm M_{\odot}$, we adopt the IMF as described by \citet{Chabrier2003}. This results in the stellar mass distribution as shown in the left panel of Figure \ref{fig:gas_mass_and_M*}. We find that the total stellar mass ($\geq 1~{\rm M_{\odot}}$) of this DLA is $\log_{10}(M_{\star}/{\rm M_{\odot}}) =4.8\pm 1.3$. We can also infer the total gas mass of the DLA using our enrichment model, as follows. Assuming that the DLA towards J0035$-$0918 has retained 100 per cent of the metals produced \emph{in-situ}, we can calculate the total gas mass required to produce the observed metal abundance. For this calculation we use the observed [O/H] abundance as a proxy of the metal abundance. The resulting distribution is shown in the right panel of Figure \ref{fig:gas_mass_and_M*}. If this is a Population III enriched system, to achieve the measured abundance [O/H] = $-2.45 \pm 0.06$, we would require a total gas mass $\log_{10}(M_{\rm gas}/{\rm M_{\odot}}) = 6.3^{+1.4}_{-0.9}$. If, in fact, some metals were not retained by the DLA, the observed [O/H] abundance could be achieved through metals mixing with a smaller reservoir of hydrogen. In this case, our inference would correspond to an upper limit of the total gas mass. We find that our inferences of the stellar and gas mass of this DLA are consistent (i.e. within $1\sigma$) of the corresponding values quoted by \citet{Welsh2019} for typical very metal-poor DLAs.
\begin{figure*}
    \centering
    \includegraphics[width=\textwidth]{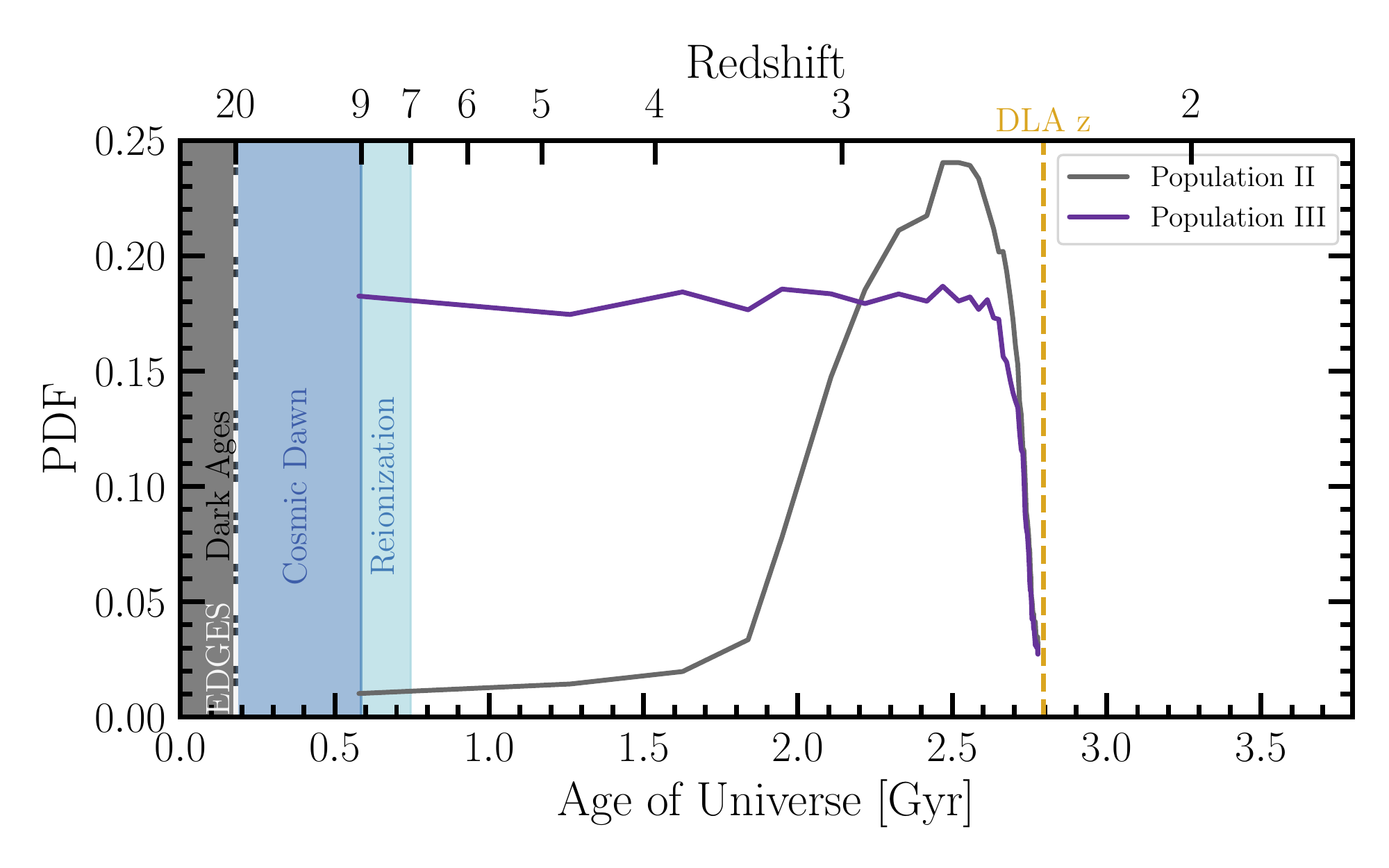}
    \caption{The most likely epoch of star formation experienced by the DLA towards J0035$-$0918 given our maximum likelihood enrichment model. The purple (Population III) and grey (Population II) curves indicates the probability that the stars which chemically enriched this system were born at a given redshift. A high value of the PDF indicates the most likely redshift that the DLA experienced a burst of star formation (or, equivalently, the epoch when most of the enriching stars were formed). We highlight the epoch of reionization as determined by the \protect\citet{Planck2018} with a light-blue shaded band. Similarly, the period known as the Cosmic Dawn is shown in dark blue. Between this epoch and the Dark Ages (black shaded band) is the recent EDGES detection (vertical white dot-dashed line) from \protect\citet{Bowman2019}.}
    \label{fig:cosmic_evolution}
\end{figure*}
\subsection{Enrichment timescale: Evidence of reionization quenching?}
\begin{figure}
    \centering
    \includegraphics[width=\columnwidth]{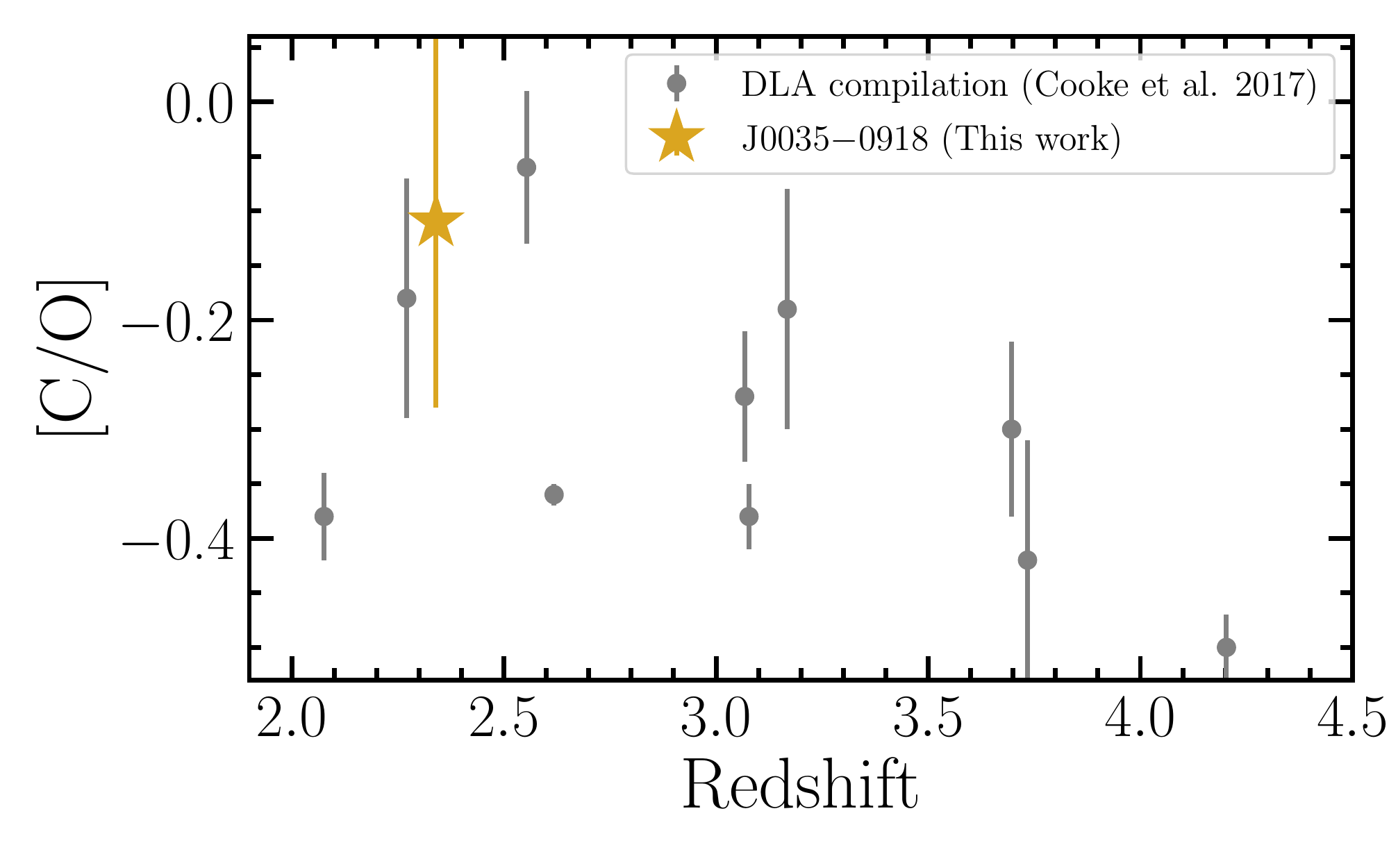}
    \caption{The redshift evolution of the measured [C/O] ratio of near-pristine DLAs. These systems are all bona fide DLAs (i.e. log$_{10}$~$N$(H\,{\sc i})/cm$^{-2}>20.3$) with metallicity [O/H]$<-1.75$ that have been observed with a high resolution (R~$>30,000$) echelle spectrograph. The potential upward trend in [C/O] at lower redshift  is supported by the [C/O] determination of J0035$-$0918.}
    \label{fig:CO_v_redshift}
\end{figure}

Our constraints on the minimum mass of the enriching stars can be used to estimate the epoch when the DLA experienced most of its star formation. Using the stellar lifetimes from \citet{Woosley2002} and \citet{Karakas2014} as well as the posterior distribution of $M_{\rm min}$ (see the histogram on the third row of Figure \ref{fig:corner_plot}), we can convert the $M_{\rm min}$ distribution to a distribution of enrichment timescales. The results of this transformation for both Population II and Population III stars are shown in Figure \ref{fig:cosmic_evolution}. Since our analysis disfavours a large ( $\sim8~\rm M_{\odot}$) minimum enriching mass, the sharp fall of these distributions as we approach the redshift of the DLA suggests that the majority of star formation must have ended prior to this epoch. We can see from this figure that if Population II stars are the predominant enrichers, then the DLA must have had a burst of star formation just a few hundred Myr before we observe the DLA today. Such a timescale is relatively short given that the DLA would need to have recovered quickly from the putative supernova feedback in order to be observed at $z=2.34$ with a significant quantity of neutral gas and apparently no turbulence. Hydrodynamic models of the enrichment of ultra-faint dwarf (UFD) galaxies \citep{Webster2015a} indicate that the chemistry of these systems requires periods of extended star formation, which may also be required to explain the enrichment of this DLA (see also, \citealt{Webster2015b}). \\
\indent In this DLA, however, we find no evidence of enrichment by low mass Population II stars. As shown in Figure \ref{fig:cosmic_evolution} by the grey curve, the most likely explanation in this scenario is that the DLA experienced no significant star formation post-reionization, for at least $\gtrsim1~{\rm Gyr}$. There are several mechanisms that can temporarily quench a low mass galaxy. One such possibility is reionization quenching (e.g. \citealt{Bullock2000}) due to the cosmic reionization of hydrogen at $z\simeq8$ (\citealt{Planck2018}; light blue band in Figure \ref{fig:cosmic_evolution}). Reionization played two critical roles that affected star formation in low mass galaxies. First, reionization heated up the intergalactic medium, thereby limiting the accretion of gas onto low mass galaxies. This starved low mass galaxies of the gas supply needed to form stars. Moreover, reionization heated up the interstellar medium of low mass galaxies, bringing a halt to any ongoing star formation.\\
\indent The main thrust of current observational efforts to study the reionization quenching of low mass galaxies utilise deep observations of the lowest mass UFD galaxies orbiting the Milky Way \citep{Brown2014, Weisz2014}. These studies use color magnitude diagrams to reconstruct the star formation histories of the UFD galaxies. This technique allows us to study the present properties of UFD galaxies in detail, but currently suffers from poor time resolution at $z\gtrsim2$. Therefore it is difficult to study the finer details of the reionization quenching process, such as the duration of the quenching and the properties of the gas that survives reionization. Simulations of low mass galaxy formation (e.g. \citealt{Wheeler2015, Onorbe2015}) indicate that reionization can bring a halt to star formation for $\sim1~{\rm Gyr}$; furthermore, these simulations indicate that some low mass galaxies are able to retain a small reservoir of gas for future star formation. \\
\indent Studying the chemical enrichment of the most metal-poor DLAs may therefore offer a novel and exciting opportunity to study reionization quenching in detail by using certain chemical tracers as a `chemical clock'. Taken at face value, our observations combined with our stochastic chemical enrichment model tentatively suggest that star formation in the DLA towards J0035$-$0918 may have experienced a $\sim1~{\rm Gyr}$ hiatus. In principle, one might be able to tease out the signature of reionization quenching by studying a sample of metal-poor DLAs; reionization is a cosmic event that comparably affects all galaxies at a given mass scale. This signature may be encoded in the star formation histories (and therefore chemistry) of the most metal-poor DLAs. One prediction of this scenario is that the most metal-poor DLAs should exhibit a general increase of their [C/O] at redshift $z\sim3$; oxygen is primarily produced by massive stars on short timescales, while carbon is produced by massive stars as well as low and intermediate mass stars on longer timescales \citep{Akerman2004, Cescutti2009, Romano2010}. Thus, an increase of [C/O] at low redshift would mark the yields of the first low and intermediate mass Population II stars to have formed post-reionization. We may be witnessing the first tentative evidence of this effect in Figure \ref{fig:CO_v_redshift} where we plot the [C/O] abundances of the \emph{most} metal-poor DLAs currently known as a function of their redshift. These data are based on the list compiled by \citet{Cooke2017} who investigated the chemical evolution of DLAs with [O/H]$<-1.75$. We consider only those with log$_{10}$~$N$(H\,{\sc i})/cm$^{-2}>20.3$ that have also been observed with a high resolution (R~$>30,000$) spectrograph. Figure \ref{fig:CO_v_redshift} shows tentative evidence that some near-pristine DLAs display slightly elevated [C/O] at lower redshift. We note that the size of the errors associated with each system is due to the relative saturation of the lines used to determine the abundance ratio. 
\subsection{Fine-structure constant}
\indent Given the quiescence and intrinsically simple cloud structure of
the DLA towards J0035$-$0918, in combination with the wavelength
stability of ESPRESSO, we have an ideal dataset for placing a
bound on the invariance of the fine-structure constant $\alpha$ at high redshift. Given that this DLA is a near-pristine environment which is presumably living in a relatively underdense part of the Universe compared to other absorption line systems, it also offers an alternative environment to test the invariance of the fundamental couplings.\\
\indent Astrophysical determinations of the variability of the fine-structure constant, $\alpha\equiv e^{2}/4\pi\epsilon_{0}\hbar c\approx1/137$, which characterises the strength of the electromagnetic force, has been the subject of many investigations since the advent of the 10~m class telescopes. The general principle is to measure the change of the fine-structure constant measured today ($\alpha_{0}$) relative to a measurement at high redshift ($\alpha_{\rm z}$), leading to a bound of the form:
\begin{equation}
\Delta\alpha/\alpha\equiv(\alpha_{\rm z}-\alpha_{0})/\alpha_{0}
\end{equation}
To obtain a measure of $\alpha_{\rm z}$, the observed wavelengths of
several spectral lines need to be measured very accurately, as each absorption line exhibits a different sensitivity to $\alpha$. This sensitivity can be parameterised by a change to the wavenumber
of a given transition at high redshift, $\omega_{\rm z}$, relative to
the same value measured in the laboratory today, $\omega_{0}$:
\begin{equation}
\omega_{\rm z} = \omega_{0} + q\,x
\end{equation}
where $q$ is the sensitivity coefficient which determines how
sensitive a given transition is to changes in $\alpha$, and
$x=(\alpha_{\rm z}/\alpha_{0})^{2} - 1$. In this work, we use
the $q$-coefficients compiled by \citet{MurphyBerengut2014}.\\
\indent For this test, we use only the ESPRESSO data and include the absorption lines of the O\,{\sc i} $\lambda$1302, Al\,{\sc ii} $\lambda$1670, Si\,{\sc ii} $\lambda$1536, Fe\,{\sc ii} $\lambda$1608, and Fe\,{\sc ii} $\lambda$2344, which exhibit sensitivity coefficients in the range $-1165\le q \le 1375$. Based on just the one absorption line system that we report here, we infer a bound on the invariance of the fine-structure constant, $\Delta \alpha$/$\alpha =(-1.2 \pm 1.1)\times10^{-5}$. Given the relatively low S/N of our data owing to the faint background quasar and short integration time, this bound is impressively tight, falling just a factor of $\sim5$ short of the precision achieved by \citet{King2012} who assembled and analysed a sample of almost 300 absorption line systems for the purpose of testing the invariance of $\alpha$ over cosmic time. In addition, the simplicity and quiescence of the cloud structure provides us with confidence that the modelling of the line profile has not introduced unaccounted for systematic uncertainties. This is the first result that we are aware of that demonstrates the superior wavelength accuracy delivered by the ESPRESSO instrument in 4UT mode.

\section{Conclusions}
\label{sec:con}
We report the first bound on the $\rm ^{12}C/^{13}C$ abundance ratio of a near-pristine environment using science verification data acquired with ESPRESSO in 4UT mode. Our main conclusions are as follows: 
\begin{enumerate}
    \item We have demonstrated that the wavelength accuracy afforded by ESPRESSO permits a limit on the $^{12}$C/$^{13}$C isotope ratio in the quiescent DLA towards J0035$-$0918 using the C\,{\sc ii}\,$\lambda$1334 absorption line. A significant quantity of $^{13}$C, if found, could be a signature of low mass metal-free star formation.
    \item We find that the gas cloud is well-modelled by a single absorption component whose broadening is entirely dominated by the thermal motions of the gas. On the basis of this model, we report a conservative $2\sigma$ lower limit $\rm log_{10}\,^{12}C/^{13}C >+ 0.37$. We therefore conclude that this DLA predominantly contains $^{12}$C. Given this 2$\sigma$ limit, we are unable to confidently rule out the presence of low mass Population III stars at this stage. 
    \item We developed a stochastic chemical enrichment model to test whether the chemistry of this system is better modelled by Population III or Population II enrichment. We have found, given current data, that both scenarios are plausible and are equally capable of producing the observed abundances of $\rm ^{12}C/^{13}C$, [C/O], [Si/O], and [Fe/O]. 
    \item  Based on our best-fitting enrichment model, we estimate the DLA contains a stellar mass of  $\log_{10}(M_{\star}/{\rm M_{\odot}}) =4.8\pm 1.3$ and a gas mass of $\log_{10}(M_{\rm gas}/{\rm M_{\odot}}) = 6.3^{+1.4}_{-0.9}$.
    \item We report tentative evidence that the \emph{most} metal-poor DLA population exhibits somewhat higher [C/O] values at redshift $z\lesssim3$. The elevated [C/O] ratios at $z\lesssim3$ might be a signature of enrichment from the first metal-enriched low and intermediate mass stars.
    \item Our enrichment model also suggests that --- if this gas cloud is predominantly enriched by Population II stars --- the bulk of the metals were produced just a few hundred Myr before the time that we observe the DLA. Prior to that, star formation in this DLA appears to have experienced a period of quiescence. We propose that this quiescence may have been caused by the cosmic reionization of hydrogen, but this can only be confirmed with future observations of near-pristine DLAs covering the redshift interval $z\simeq 2-4$.
    \item We use the simplicity of the absorption profile of this system to investigate whether there is a detectable spatial or temporal variation of the fine-structure constant. When including $\Delta\alpha$/$\alpha$ as a free parameter in our line-fitting procedure, we find $\Delta \alpha$/$\alpha = (-1.2 \pm 1.1)\times10^{-5}$.
\end{enumerate}
Our work demonstrates the wealth of information made available through studying the chemistry of near-pristine absorption systems. The detailed abundance patterns of the most metal-poor DLAs provide insight into the earliest episodes of chemical enrichment. Indeed, this first bound on the C isotope ratio has ruled out the presence of strong $^{13}$C in this DLA; with data of S/N=20, we could observationally rule out significant enrichment by low mass Population III stars in this near-pristine environment. A similar study across the metal-poor DLA population would determine whether these systems typically show signatures of Population II or Population III enrichment. From these reconstructed enrichment histories, we may find observational evidence of reionization quenching at $2<z<4$ (within the redshift interval where these absorption systems are most easily studied), and be able to study the physical properties (e.g. density, temperature) of the gas affected by reionization quenching. Thanks to ESPRESSO, studies of this nature are now within the realm of possibility.

\section*{Acknowledgements}
We are grateful to the staff astronomers at the VLT for their assistance with the observations.
During this work, R.~J.~C. was supported by a
Royal Society University Research Fellowship.
We acknowledge support from STFC (ST/L00075X/1, ST/P000541/1).
This project has received funding from the European Research Council 
(ERC) under the European Union's Horizon 2020 research and innovation 
programme (grant agreement No 757535).
This work used the DiRAC Data Centric system at Durham University,
operated by the Institute for Computational Cosmology on behalf of the
STFC DiRAC HPC Facility (www.dirac.ac.uk). This equipment was funded
by BIS National E-infrastructure capital grant ST/K00042X/1, STFC capital
grant ST/H008519/1, and STFC DiRAC Operations grant ST/K003267/1
and Durham University. DiRAC is part of the National E-Infrastructure.
This research has made use of NASA's Astrophysics Data System, and the following software packages:
Astropy \citep{ASTROPY},
Corner \citep{CORNER},
Matplotlib \citep{MATPLOTLIB},
and NumPy \citep{NUMPY}.



\bibliographystyle{mnras}
\bibliography{references} 



\appendix


\bsp	
\label{lastpage}
\end{document}